\documentclass[aps,groupedaddress,superscriptaddress,amsmath,amssymb,twocolumn,pra,longbibliography]{revtex4-1}
\usepackage[dvips]{graphicx}
\usepackage{stmaryrd}
\usepackage{color}
\usepackage[normalem]{ulem}
\usepackage{bm}
\usepackage{amsmath}
\usepackage{amssymb}

\def\pb{{\bf p}}
\def\qb{{\bf q}}

\def\n{{\bm\nabla}}
\def\kb{{\bf k}}

\begin{document}
\author{Yukihiro Tadokoro}
\affiliation{Toyota Central R\&D Labs., Inc., Nagakute, Aichi, 480-1192, Japan}
\author{Hiroya Tanaka}
\affiliation{Toyota Central R\&D Labs., Inc., Nagakute, Aichi, 480-1192, Japan}
\author{M. I. Dykman}
\affiliation{Department of Physics and Astronomy, Michigan State  University, East Lansing, MI 48824, USA}

\title{Noise-induced switching from a symmetry-protected shallow metastable state}

\date{\today}

\begin{abstract}
We consider escape from a metastable state of a nonlinear oscillator driven close to triple its eigenfrequency. The oscillator can have three stable states of period-3 vibrations and a zero-amplitude state. Because of the symmetry of period-tripling, the zero-amplitude state remains stable as the driving increases. However, it becomes shallow in the sense that the rate of escape from this state exponentially increases, while the system still lacks detailed balance. We find the escape rate and show how it scales with the parameters of the oscillator and the driving. The results facilitate using nanomechanical, Josephson-junction based, and other mesoscopic vibrational systems for studying, in a well-controlled setting, the rates of rare events in systems lacking detailed balance. They also describe how fluctuations spontaneously break the time-translation symmetry of a driven oscillator.
\end{abstract}

\maketitle
%
%

\section{Introduction}

Fluctuation-induced switching from a metastable state underlies a broad range of phenomena and has been attracting much interest in diverse areas, from statistical physics to chemical kinetics, biophysics, and populatio'n dynamics. Where fluctuations are weak on average, switching is a rare event, with the rate much smaller than the relaxation rate of the system. For classical and quantum systems in thermal equilibrium, switching has been well understood \cite{Kramers1940,Leggett1987,Kagan1992}. Here the major mechanisms of switching are thermal activation over the free energy barrier or, for low temperatures, tunneling. The corresponding theory has been standardly used to characterize Josephson junctions  \cite{Kurkijarvi1972,Fulton1974,Kautz1996}, to study magnetic systems \cite{Brown1963,Wernsdorfer1997,Garanin1997a,Coffey1998a,Ingvarsson2000}, and for other applications.

Much progress has been made  over the last few decades on the theory of switching in systems far from thermal equilibrium, see Refs.~\cite{Graham1986a,Freidlin1998,Luchinsky1998,Touchette2009,Kamenev2011,Bertini2015,Assaf2017} for a review.  However, many problems remain open on the theory side, and much remains to be learned on the  experimental side. 

The experiments require well characterized nonequilibrium systems that remain stable for a time much longer than the relaxation time. A class of systems that stand out in this respect are resonantly driven mesoscopic vibrational systems where switching occurs between the states of forced vibrations. The examples range from electrons in a Penning trap to cold atoms to nano- and micromechanical systems to Josephson junction based systems, cf. \cite{Lapidus1999,Siddiqi2004,Aldridge2005,Kim2005,Stambaugh2006,Chan2007,Chan2008a,Vijay2009,Heo2010,Wilson2010,Dykman2012,Venstra2013,Defoort2015,Dolleman2019,Andersen2019}.

A most detailed comparison of the theory and the experiment can be done for ``shallow'' metastable states. These are states with a comparatively low barrier for escape. A simple example is a state at the bottom of a shallow potential well. Even a comparatively weak noise can lead to escape from a shallow state with an appreciable rate. This significantly simplifies the experiment. Typically, a stable state becomes ``shallow'' when one of the parameters of the system approaches the value  where  the state disappears (a bifurcation point). The rate of switching from a shallow state displays a characteristic scaling with the distance to the bifurcation point in the parameter space. For stable states of forced vibrations such scaling has been found in the classical and quantum regimes \cite{Dykman1979b,Dykman1998,Dykman2007} and has been observed both for the vibrations at the frequency of the driving resonant field \cite{Aldridge2005,Stambaugh2006,Vijay2009,Defoort2015,Dolleman2019} and for parametrically excited vibrations at half the drive frequency \cite{Kim2005,Chan2007}. 

A major feature that underlies the scaling is that, even though the stable states are vibrational, the problem can be mapped on fluctuations of an overdamped particle in a one-dimensional potential well. This is a consequence of the onset of a ``soft mode'' that controls the motion near the relevant bifurcation points \cite{Guckenheimer1997}. Moreover, because there is only one slow variable, the corresponding slow motion has detailed  balance.

In this paper we consider escape from a shallow metastable state with no detailed balance. We show that such a state exists even in a simple system which has only two dynamical variables. The dynamics is not controlled by soft modes. Rather the emergence of the shallow state is a consequence of the symmetry of the system. The considered model is minimalistic:  the scaled equations of motion without noise contain a single parameter. 

The physical system we consider is a vibrational mode driven close to triple eigenfrequency. The classical dynamics of such a mode in the absence of noise has been well understood \cite{Nayfeh2004}. For sufficiently strong driving, the mode can have three stable period-3 vibrational states, which all have the same amplitude and differ in phase by $2\pi/3$. However, the state with no vibrations (except maybe vibrations at the drive frequency with a very small amplitude) is also stable. It remains stable as the driving amplitude increases. We call it a zero-amplitude state. 
\begin{figure}[t]
\includegraphics[width=3.6cm]{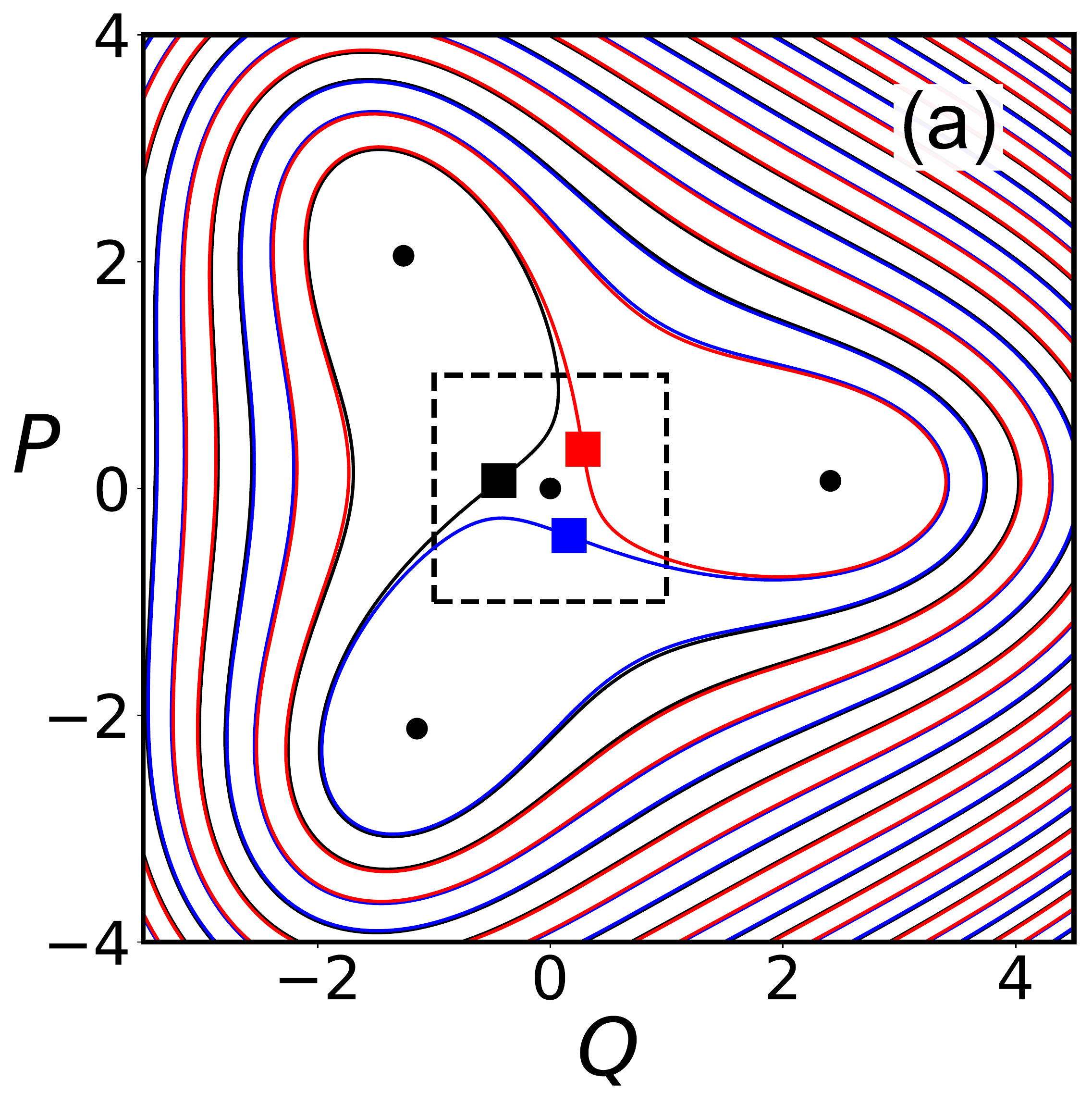} \hfill
\includegraphics[width=3.8cm]{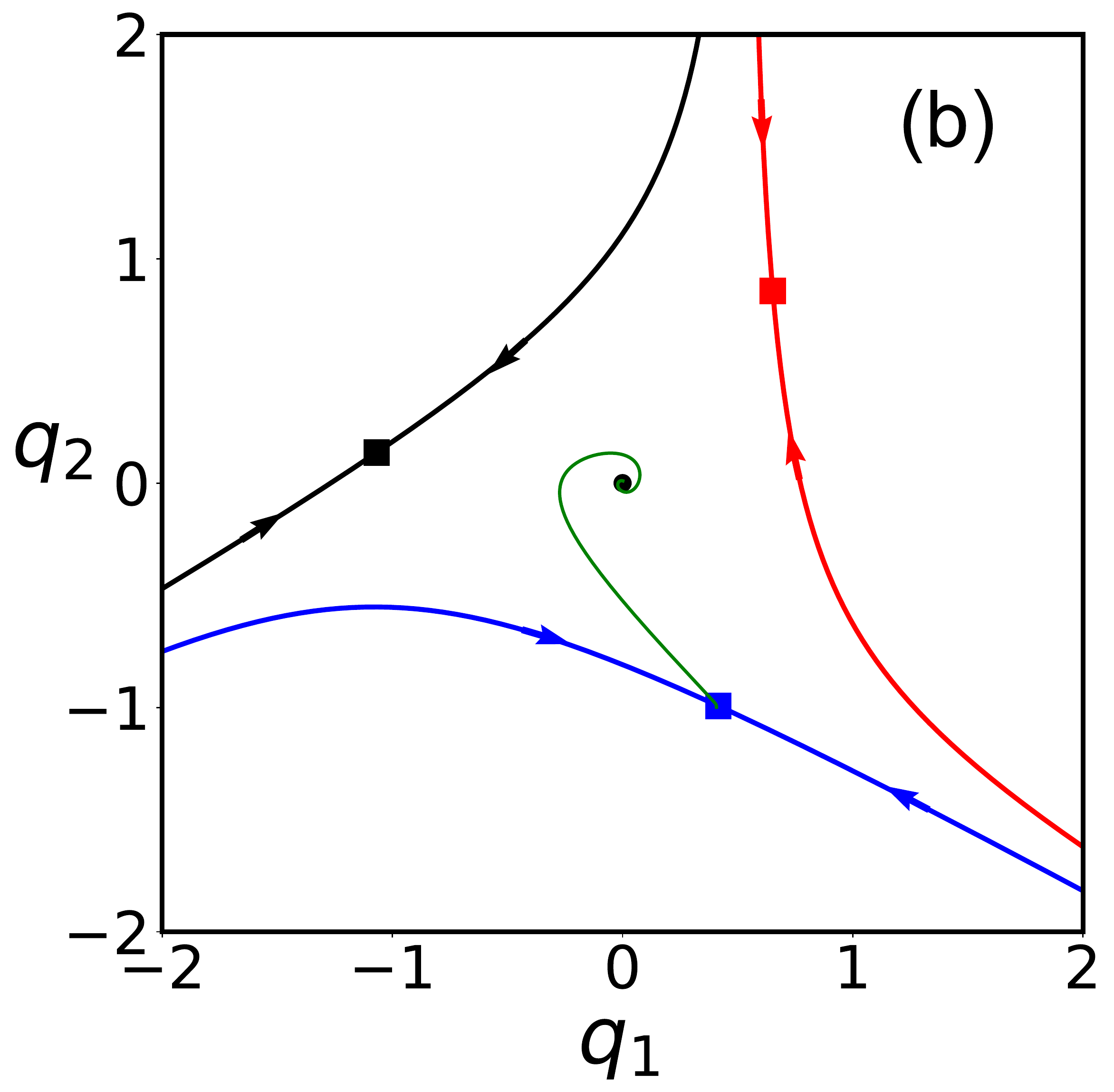}\\
\includegraphics[scale=0.35]{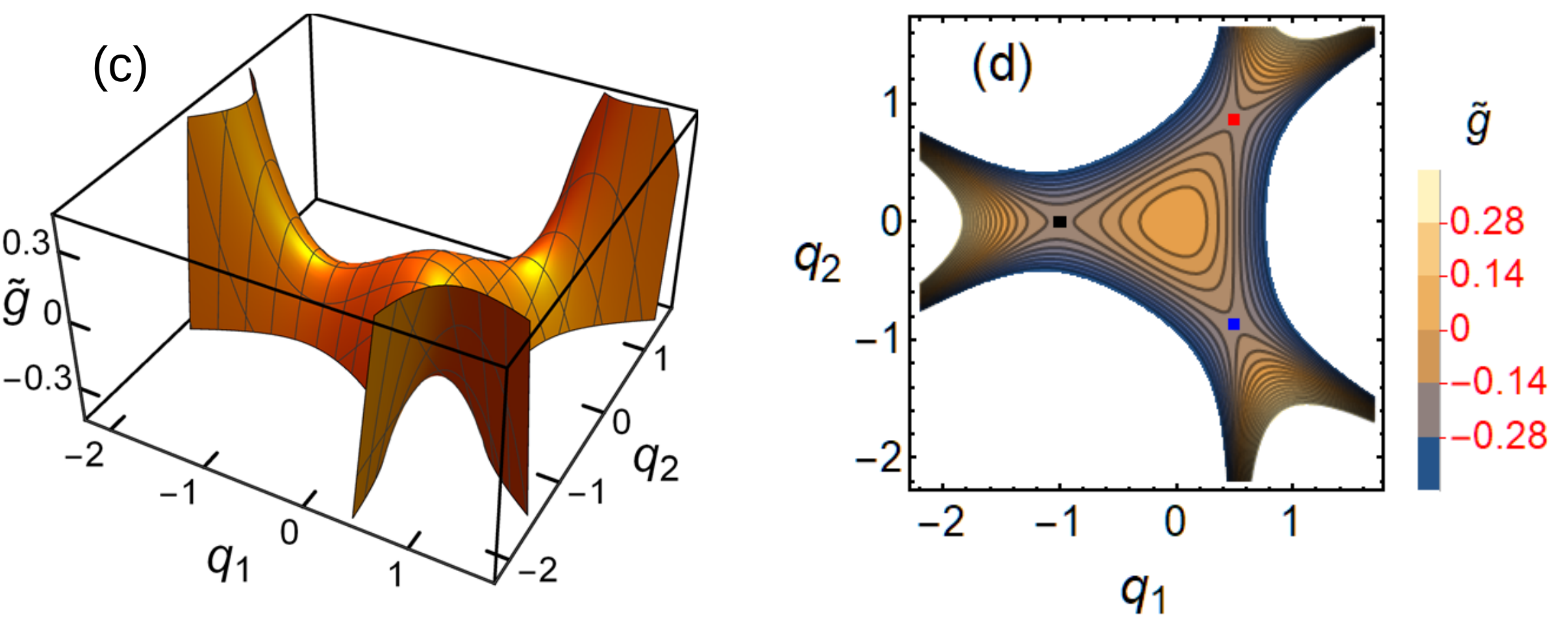}
\caption{(a) The phase portrait of the mode that displays period tripling; $Q$ and $P$ are the quadratures (the coordinate and momentum in the rotating frame). Circles and squares show the stable states and the saddle points, respectively.  The lines show the separatrices.  The plot refers to $\kappa= 0.4$ and $f=2$ in Eq.~(\ref{eq:eom}). (b) The part of the phase portrait inside the dashed square in (a) in the scaled coordinate and momentum $q_1 = fQ$  and $q_2=fP$. For $f\gg 1$ the dynamics is described by Eq.~(\ref{eq:eom_reduced}). The green line that comes from the stable state shows the most probable path followed in escape for $\kappa =0.4$. (c) The effective Hamiltonian (\ref{eq:tilde_g}) of motion around the shallow state  $q_1=q_2=0$ in the absence of dissipation. With dissipation, the local maximum at the origin becomes a stable state. (d) The contour plot of the Hamiltonian (c). The squares show the saddle points. These points shift in the presence of dissipation, as seen in (b), but the stable state remains at $q_1=q_2=0$.}
\label{fig:phase_portrait}
\end{figure}

The basin of attraction of the zero-amplitude state, i.e., the range in which the mode approaches this state from an initially prepared state,can be seen from Fig.~\ref{fig:phase_portrait} (a) and (b). It is the interior of the region centered at the origin and limited by the solid lines (the  separatrices). As we show, the size of this region decreases with the increasing driving. However, it remains nonzero for a finite driving amplitude. Indeed, annihilation of the zero-amplitude state would require that it merges simultaneously with the three saddle points, which correspond to unstable period-3 vibrations and are shown by squares in Fig.~\ref{fig:phase_portrait}. The persistence of the zero-amplitude state is therefore a consequence of the symmetry, which is the symmetry of the period-3 vibrational states with respect to incrementing their phase by $2\pi/3$ or, in other words, the symmetry of Fig.~\ref{fig:phase_portrait} with respect to rotation by $2\pi/3$. 

As the basin of attraction of the zero-amplitude state shrinks with the increasing driving, this state becomes more and more shallow. Thus, for strong driving,  the zero-amplitude state is a symmetry-protected shallow metastable state. We are not aware of a simpler shallow metastable state with the dynamics that is not controlled by a soft mode no matter how small the basin of attraction becomes.

Resonant period tripling leads to unusual quantum dynamics of a vibrational mode \cite{Guo2013a,Zhang2017c,Zhang2019,Gosner2019}. In the quantum regime, a period tripling was recently observed in a  superconducting resonator with several nonlinearly coupled modes \cite{Svensson2017a,Svensson2018}. There is a significant difference between quantum fluctuations about the shallow zero-amplitude state and the shallow states near bifurcation points.  Quantum fluctuations near a bifurcation point are very similar to classical fluctuations, since the dynamics is controlled by a single slow variable \cite{Dykman2007}. In contrast, the dynamics near the considered shallow zero-amplitude state is described by two non-commuting dynamical variables, and therefore quantum fluctuations qualitatively differ from classical fluctuations. In the present paper we study escape due to classical fluctuations.

\section{The model}
\label{sec:model}

Mesoscopic vibrational systems, from nano- and micro-mechanical modes to Josephson-junction based systems, are nonlinear. A major effect of nonlinearity is the dependence of the vibration frequency on the amplitude. To the leading order in the nonlinearity, this dependence is well described by the Duffing model, in which one takes into account the quartic term in the expansion of the potential energy in the mode coordinate $q$. To resonantly excite period-3 vibrations one has to apply a force at frequency $\omega_F$ close to triple the eigenfrequency $\omega_0$. The simplest Hamiltonian  that describes the mode dynamics is 
\begin{align}
\label{eq:hamiltonian}
H=\frac{1}{2}(p^2 + \omega_0^2 q^2) +\frac{1}{4} \gamma q^4 -\frac{1}{3}Fq^3\cos\omega_F t.
\end{align}
Here, $p$ is the mode momentum, $\gamma$ is the nonlinearity parameter, $F$ is the driving amplitude, and we have set the mass of the mode $m=1$. The condition of the driving being resonant means that the frequency detuning $\delta\omega$ is small,
\[ |\delta\omega|\ll \omega_0, \qquad \delta\omega = \frac{1}{3}\omega_F - \omega_0.\]
While assuming the detuning small, we will further assume that it is nonzero and that both $\delta\omega$ and $\gamma$ are positive. These conditions are not necessary, the results immediately extend to the cases where $\delta\omega,\gamma<0$ and also to $\delta\omega=0$; the results extend also to the case where the driving is additive and the term in Eq.~(\ref{eq:hamiltonian}) that describes the driving has the form $-F' q \cos\omega_Ft$,  cf. \cite{Zhang2019}.  

Along with the detuning, we will assume that the nonlinearity and the driving are also small, so that the vibrations are nearly sinusoidal and, equivalently, the change of the mode frequency due to the nonlinearity is small for the typical displacement $\bar q$, i.e., $\gamma {\bar q}^2, F\bar q \ll \omega_0^2$.   

\subsection{The rotating wave approximation}
\label{subsec:RWA}

Resonant dynamics of the mode is conveniently described by the amplitude and phase of the vibrations at frequency $\omega_F/3$, which vary on the time scale that largely exceeds the vibration period $2\pi/\omega_F$. To describe this dynamics we switch to the frame that rotates at frequency $\omega_F/3$ and introduce the scaled coordinate $Q$ and momentum $P$ in this frame. The corresponding transformation  is 
\begin{align}
\label{eq:new_variables}
&q(t)+(3i/\omega_F)p(t) = (8\omega_F\,\delta\omega\,/9\gamma)^{1/2}\nonumber\\
&\qquad\times [Q(t) +iP(t)]\exp[-i\omega_Ft/3]
\end{align}
[note that these are two equations for two real variables $Q(t)$ and $P(t)$].

The transformed Hamiltonian $H$ contains time-independent terms and terms that oscillate at the frequency $2\omega_F/3$ and its overtones. The effect of  these terms is small in the considered parameter range and can be disregarded, i.e., the dynamics can be analyzed in the rotating wave approximation (RWA). In the RWA, the Hamiltonian (\ref{eq:hamiltonian}) in the variables $Q,P$ becomes $H\to [8(\omega_F\,\delta\omega)^2/27\gamma]g(Q,P)$, where
\begin{align}
\label{eq:Hamiltonian_RWA}
&g(Q,P) = \frac{1}{4}(Q^2+P^2 -1)^2 -\frac{1}{3} f(Q^3 - 3PQP), \nonumber\\
&f=F/(8\omega_F \gamma \delta\omega)^{1/2}.
\end{align}
The Hamiltonian $g$ contains a single parameter, the scaled amplitude of the driving field $f$. In the considered range of small nonlinearity and small detuning $\delta\omega$, parameter $f$ can be large even for weak driving.

The coupling of the mode to a thermal reservoir leads to decay of the vibrations and to noise. For several microscopic mechanisms of the coupling, on the time scale that largely exceeds $\omega_0^{-1}$ and the correlation time of the relevant fluctuations of the reservoir, the dynamics of the weakly nonlinear mode is Markovian, cf.~\cite{Dykman1979b}. In the simplest case of the coupling linear in the mode coordinate, the dynamical equations of motion in the dimensionless time $\tau = (\delta\omega)t$ read 
\begin{align}
\label{eq:eom}
\frac{dQ}{d\tau} = K_Q(Q,P)+ \xi_Q(\tau),\qquad &K_Q =  \partial_P g -\kappa Q  ,\nonumber\\
\frac{dP}{d\tau}=K_P(Q,P) + \xi_P(\tau),\qquad &K_P=  -\partial_Q g -\kappa P .
\end{align}
Here $\kappa$ is the dimensionless decay rate of the mode. It is simply related to the friction coefficient $\Gamma$ used when the mode dissipation is phenomenologically described by a friction force $-2\Gamma dq/dt$ and the friction is weak, $\Gamma\ll \omega_0$; in this case, $\kappa = \Gamma/\delta\omega$. However, Eq.~(\ref{eq:eom}) can apply even where in the laboratory frame the dynamics is non-Markovian. 

The terms $\xi_Q(\tau)$ and $\xi_P(\tau)$ in Eq.~(\ref{eq:eom}) are independent $\delta$-correlated Gaussian noises, 
\begin{align}
\label{eq:noise}
\langle\xi_Q(\tau)\xi_Q(0)\rangle = \langle\xi_P(\tau)\xi_P(0)\rangle = 2D\delta(\tau).
\end{align}
For thermal noise $D=81k_B T\kappa\gamma/8\omega_F^3\,\delta\omega$.

\subsection{The phase portrait}
\label{subsec:phase_portrait}

The phase portrait of the mode in the rotating frame in the absence of noise is shown in Fig.~\ref{fig:phase_portrait}(a). For $f^2>2[(1+\kappa^2)^{1/2} -1]$, Eq.~(\ref{eq:eom}) has three stable stationary solutions, which correspond to three stable period-3 vibrational states with nonzero scaled squared amplitude $Q_a^2 + P_a^2$ in the laboratory frame, three saddle points, which correspond to unstable  vibrational states with scaled squared amplitude  $Q_s^2 + P_s^2$, and a stable state with $Q=P=0$,
\begin{align}
\label{eq:equilibrium}
(Q^2 + P^2)_{a,s} = 1+\frac{1}{2}f^2 \pm \frac{1}{2}(f^4 +4f^2-4\kappa^2)^{1/2}.
\end{align}
The stable states and the saddle points with the squared amplitudes (\ref{eq:equilibrium}) are located at the vertices of equilateral triangles, cf. Fig.~\ref{fig:phase_portrait}~(b). In a standard fashion, the separatrices go through the saddle points and divide the phase plane $(Q,P)$ into the basins of attraction of different stable states.  Importantly, the basins of attraction of the states with nonzero vibrational amplitude have a common boundary only with the zero-amplitude state, but not with each other. 

With the increasing amplitude of the driving field $f$, the stable states with nonzero amplitude (\ref{eq:equilibrium}) move away from the origin. In contrast, the saddle points move toward the origin,  $(Q^2+P^2)_s \approx (1+\kappa^2)/f^2$ for large $f$. Therefore, the basin of attraction of the zero-amplitude state shrinks, but at the same time, this state remains stable for any finite $f$.


\section{Dynamics near the zero-amplitude state for large driving amplitudes}
\label{sec:large_f}

For strong driving, the dynamics of the system in the vicinity of the zero-amplitude state can be conveniently studied by switching to the variables  $q_1 = fQ$ and $q_2 = fP$. They can be thought of as components of a vector $\qb \equiv (q_1,q_2)$. For $f\gg 1$,  from Eq.~(\ref{eq:eom})  the equations of motion for $q_{1,2}$ in the range $|q_{1,2}|={\cal O}(1)$ have the form
\begin{align}
\label{eq:eom_reduced}
\frac{d\qb}{d\tau} = \kb (\qb) +{\bm\xi}(\tau), \qquad \kb = (\hat\epsilon \n)\tilde g -\kappa\qb.
\end{align}  
Here $\hat\epsilon$ is the Levi-Civita tensor, $\hat\epsilon_{ii}=0$ and $\hat\epsilon_{12}=-\hat\epsilon_{21}=1$; $\n\equiv(\partial_{q_1},\partial_{q_2})$ is the gradient, 
\begin{align}
\label{eq:tilde_g}
\tilde g(\qb) =  -\frac{1}{2}\qb^2  -\frac{1}{3}(q_1^3 - 3q_1 q_2^2), 
\end{align}
and $\xi_{1,2}(\tau)$ are two independent white Gaussian noises, 
\[\langle\xi_i(\tau)\xi_j(0)\rangle = 2\tilde D\delta_{ij}\delta(\tau), \qquad \tilde D = f^2 D.\]

Function $\tilde g(\qb)$ is an analog of a Hamiltonian, if one thinks of $q_1$ as a coordinate and $q_2$ as a momentum. It determines the conservative  motion of the strongly modulated mode in the absence of dissipation and noise.  It has no parameters. Its structure is therefore universal. It is shown in Fig.~\ref{fig:phase_portrait}~(c) and (d). As seen in this figure, $\tilde g(\qb)$ has a local maximum  at $\qb = {\bf 0}$ and three saddle points. The local maximum corresponds to the zero-amplitude state, in the presence of dissipation. 

The phase portrait of the system (\ref{eq:eom_reduced}) in the presence of dissipation but with no noise is shown in Fig.~\ref{fig:phase_portrait}~(b). Dissipation shifts the saddle points of $\tilde g(\qb)$, but they remain saddle points. We use  the same color coding for the saddle points of $\tilde g(\qb)$ in Fig.~\ref{fig:phase_portrait}~(d) and the saddle points of the full dynamics (\ref{eq:eom_reduced}) in Fig.~\ref{fig:phase_portrait}~(a). The separatrices in Fig.~\ref{fig:phase_portrait}~(b) are the boundaries of the basin of attraction to the zero-amplitude state. 

Overall, in the absence of noise the dynamics described by Eq.~(\ref{eq:eom_reduced}) depends on only one parameter, the scaled decay rate $\kappa$. Varying $\kappa$ leads only  to a quantitative change of the phase portrait, the structure remains intact, as seen from Fig.~\ref{fig:phase_portrait}~(b). The simple topology of the phase portrait indicates  the universality of  the dynamics. Quite remarkably, this universality is not related to the onset of a soft mode. There is no slowing down.

Another important feature of the dynamics (\ref{eq:eom_reduced}) is that the effective noise intensity $\tilde D$ increases with the increasing driving amplitude $F\propto f$. Such increase is essentially a consequence of the rescaling of the dynamical variables. As the basin of attraction of the zero-amplitude state on the original $(Q,P)$ phase plane shrinks with the increasing $f$, the noise becomes effectively stronger. This is what makes the zero-amplitude state ``shallow''.

By linearizing equations of motion about $\qb={\bf 0}$ we see that, for weak noise, the mean-square displacement about the zero-amplitude state is
\[\langle \qb^2\rangle = 2\tilde D/\kappa. \]
The linearization makes sense only if $\langle \qb^2\rangle \ll \qb_s^2$, where 
\[\qb_s^2 = 1+\kappa^2\]
is the squared distance of the saddle points from the origin, with the account taken of the dissipation. The positions of the three saddle points can be written as $q_{1s}+ iq_{2s}=q_s\exp(i\phi_s)$ with $\exp(3i\phi_s) =- (1-i\kappa)/q_s$. The latter equation gives three values of $\phi_s$ that correspond to different saddle points and differ by $2\pi/3$. 

With the increasing $\tilde D\propto f^2$, the mean-square displacement $\langle\qb^2\rangle$ increases. Once it approaches $\qb_s^2$, fluctuations about the stable state may no longer be assumed small. Physically, it means that the system placed initially near the zero-amplitude state quickly escapes from the basin of attraction of this state and switches to one of the states of period-3 vibrations. For smaller $f$ the escape rate is smaller, but still it may be not exceedingly small even for a weak noise, justifying the term ``shallow metastable state'' as applied to the zero-amplitude state. 

\section{Escape rate from the shallow state with no detailed balance}
\label{sec:escape_rate}

The theory of escape from a metastable state of a white-noise driven system is well-established \cite{Freidlin1998}. The underlying idea is that, when the noise is weak on average, escape occurs as a result of a rare fluctuation. In this fluctuation, a large outburst of noise drives the system over the boundary of the basin of attraction of the initially occupied stable state. Once the boundary is crossed, noise is no longer needed, the system moves to another state ``on its own''. An outburst of noise is a certain time evolution of a random force. The outbursts needed for switching are exponentially unlikely for a weak Gaussian noise. In addition, the probabilities of different appropriate outbursts are exponentially different. The rate of escape is determined by the most probable of them, i.e., by the least improbable appropriate evolution of the random force in time. Through the equations of motion, such force leads to the corresponding trajectory of the system \cite{Feynman1965}. This trajectory is often called \cite{Maier1993a,Kautz1996} the most probable escape path (MPEP).

We will consider escape from the zero-amplitude state using Eq.~(\ref{eq:eom_reduced}). The conventional analysis refers to systems with no symmetry. In contrast, our system has a three-fold symmetry. The attraction basin of the zero-amplitude state  is bound by three separatrices, which can be obtained from each other by a rotation by $2\pi/3$ on the $(q_1,q_2)$ plane, see Fig.~\ref{fig:phase_portrait}~(b). The probability to cross any of them in escape is the same, and therefore the total escape rate is three times the escape rate for crossing one of them.

For weak noise, it is most probable to cross a separatrix near the saddle point \cite{Dykman1979b,Maier1997,Luchinsky1999}. For the system (\ref{eq:eom_reduced}), the rate of escape over one of the separatrices $W_0$ for small noise intensity  has the form
\begin{align}
\label{eq:escape_general}
W_0 = C\exp(-R/\tilde D), \qquad \tilde D\ll 1,
\end{align}
where  $C$ is a constant that smoothly (nonexponentially) depends on the parameters. Equation (\ref{eq:escape_general}) reminds the Kramers formula \cite{Kramers1940} for the rate of activated escape from a potential well. However, the dynamics (\ref{eq:eom_reduced}) does not correspond to Brownian motion in a potential well, and  $R$ is not a height of a potential barrier. 

The effective activation energy $R$ for the considered fluctuating dissipative system is given by the action of an auxiliary conservative system  \cite{Freidlin1998},
\begin{align}
\label{eq:R_general}
&R=\frac{1}{2}\min \int_{-\infty}^\infty d\tau {\cal  L} \left(\frac{d\qb}{d\tau},\qb\right),\nonumber\\
 & {\cal L}\left(\frac{d\qb}{d\tau},\qb\right) = \frac{1}{2}\left[\frac{d \qb}{d\tau}  - \kb(\qb)\right]^2.
 \end{align}
Here ${\cal L}$ is the Lagrangian of an auxiliary system; this is a Hamiltonian system, with no relaxation and no noise. Its appropriate Hamiltonian trajectory $\qb_{\rm opt}(\tau)$ gives the MPEP , i.e., the trajectory which the initial dissipative noisy system is most likely to follow in escape. This trajectory starts at $\qb={\bf 0}$ for $\tau \to -\infty$ and goes to one of the saddle points for $\tau\to\infty$ , see Appendix A. 

\subsection{The effective activation energy in the limiting cases}
\label{subsec:explicit}

Equation (\ref{eq:R_general}) allows one to calculate the effective activation energy $R$ for any $\kappa$ by numerically solving the variational equations of motion, or the corresponding Hamiltonian equations, with the appropriate boundary conditions, see Appendix A.
Analytically, explicit values of $R$ can be calculated in the limiting cases of large and small $\kappa$. We start with the case $\kappa\ll 1$. For $\kappa=0$  the noise-free trajectories of the system (\ref{eq:eom_reduced}) are closed loops with $\tilde g(\qb)={\rm const}$, as seen in Fig.~\ref{fig:phase_portrait}~(d). For small $\kappa$ the noise-free trajectories become tight spirals that spiral toward $\qb = {\bf 0}$, with $\tilde g(\qb)$ slowly increasing toward $\tilde g({\bf 0}) = 0$. The MPEP $\qb_{\rm opt}(\tau)$ is also a tight spiral, but it spirals from $\qb = {\bf 0}$ toward the saddle points. To the lowest order in $\kappa$, the value of $R$ is determined by the action (\ref{eq:R_general}) accumulated on this trajectory until $\tilde g$ reaches its value at the saddle point  \cite{Dykman1979b,Dmitriev1986,Chinarov1993}. 

Importantly, the multiplicity of the saddle points makes no difference, to the leading order in $\kappa$. Indeed, for $\kappa\to 0$ all saddle points have the same $\tilde g_s\equiv \tilde g(\qb_s) = -1/6$, cf. Fig.~\ref{fig:phase_portrait}~(c) and (d). Therefore the variational equations for the optimal path can be solved in the same way as in systems with a single saddle point. The resulting general expression for $R$ is \cite{Dykman1979b,Dmitriev1986,Chinarov1993}
\begin{align}
\label{eq:small_damping}
&R\approx
 \kappa \int_0^{\tilde g_s} d\tilde g \frac{M(\tilde g)}{N(\tilde g)}, \qquad M(\tilde g) = \iint_{{\cal A}(\tilde g)} dq_1 dq_2,\nonumber\\
&N(\tilde g) = \frac{1}{2}\iint_{{\cal A}(\tilde g)} \n^2 \tilde g(\qb)\,dq_1 dq_2  \qquad (\kappa\ll 1),
\end{align}
where ${\cal A}(\tilde g)$ is the interior of the orbit $dq_1/d\tau = \partial \tilde g/\partial q_2, \; dq_2/d\tau = -\partial \tilde g/\partial q_1$ with a given $\tilde g$. Such orbits are shown in Fig.~\ref{fig:phase_portrait}~(d). From Eq.~(\ref{eq:eom_reduced}), the orbit is a circle $\qb^2 = -2\tilde g$ for small $|\tilde g|$ and becomes a triangle for $\tilde g = \tilde g_s = -1/6$, with the two sides given by $q_2 = \pm (q_1+1)/\sqrt{3}$ and the 3rd side given by $q_1 = 1/2$. 

From Eq.~(\ref{eq:eom_reduced}), $\n^2 \tilde g = -2$, and therefore 
\begin{align}
\label{eq:R_underdamped_explicit}
R = \kappa/6,\qquad \kappa\ll 1.
\end{align}
This expression shows that the effective activation energy $R$ linearly increases with the scaled decay rate $\kappa$ where $\kappa$ is small.

We now consider the case of fast decay, $\kappa\gg 1$. Still we assume that $\kappa\ll f$, so that the dynamics in the considered part of the phase plane is described by Eq.~(\ref{eq:eom_reduced}).  In this case it is convenient to change variables in Eqs.~(\ref{eq:eom_reduced}) and (\ref{eq:R_general}) by setting $\qb = \kappa\qb', \tau = \tau'/\kappa$. Then $d\qb/d\tau = \kappa^2 d\qb'/d\tau'$, whereas, to the leading order in $\kappa$, the components of the vector $\kb$ become $\kb(\qb) = \kappa^2\kb'(\qb')$ with $k_1' = -q_1' + 2q_1'q_2'$, and $k_2' = -q_2' + q_1'{}^2 - q_2'{}^2$. Therefore we can write the expression for the activation energy as 
\begin{align}
\label{eq:large_kappa}
&R=\kappa^3 R',\qquad R'=\frac{1}{2}\min\int_{-\infty}^\infty d\tau' {\cal L}'\left(\frac{d\qb'}{d\tau'},\qb'\right),\nonumber\\
&{\cal L}'\left(\frac{d\qb'}{d\tau'},\qb'\right) = \frac{1}{2}\left[\frac{d\qb'}{d\tau'} -\kb'(\qb')\right]^2,\qquad \kappa\gg 1.
\end{align}

The parameter $\kappa$ has been scaled out of Eq.~(\ref{eq:large_kappa}). The Lagrangian ${\cal L}'$ does not contain any parameters. Therefore $R'$ is a number. This number can be found by solving the variational problem (\ref{eq:large_kappa}) numerically for the extreme trajectories that start at $\qb'={\bf 0}$ at $\tau'\to -\infty$ and for $\tau\to \infty$ go to one of the saddle points, which are located at $(q_1',q_2') = (0,-1), (\pm \sqrt{3}/2,1/2)$, see Appendix A. The numerical solution gives $R'\approx 0.171$. Therefore, %
\begin{align}
\label{eq:R_large_kappa}
R\approx 0.171 \,\kappa^3, \qquad \kappa\gg 1.
\end{align}

\section{Numerical results}
\label{sec:numerical}

We have studied escape from the zero-amplitude state numerically using three approaches. First, we performed numerical simulations of the full stochastic equations of motion (\ref{eq:eom}) for several values of the scaled driving force $f$. Then we performed simulations of the scaled stochastic equations (\ref{eq:eom_reduced}) that refer to large $f$ and do not contain $f$ other than in the noise intensity $\tilde D = f^2 D$. Then we solved the noise-free variational problem (\ref{eq:R_general}). The numerical integration of the stochastic equations was done following the standard routine, cf.~\cite{Mannella2002}. To find the escape rate $W_0$, for each parameter value we assembled 3000 trajectories that went from the vicinity of the zero-amplitude state to the area well behind the basin of attraction of this state. We made sure that the result was independent of the chosen boundary of this area.

\begin{figure}[t]
\includegraphics[scale=0.27]{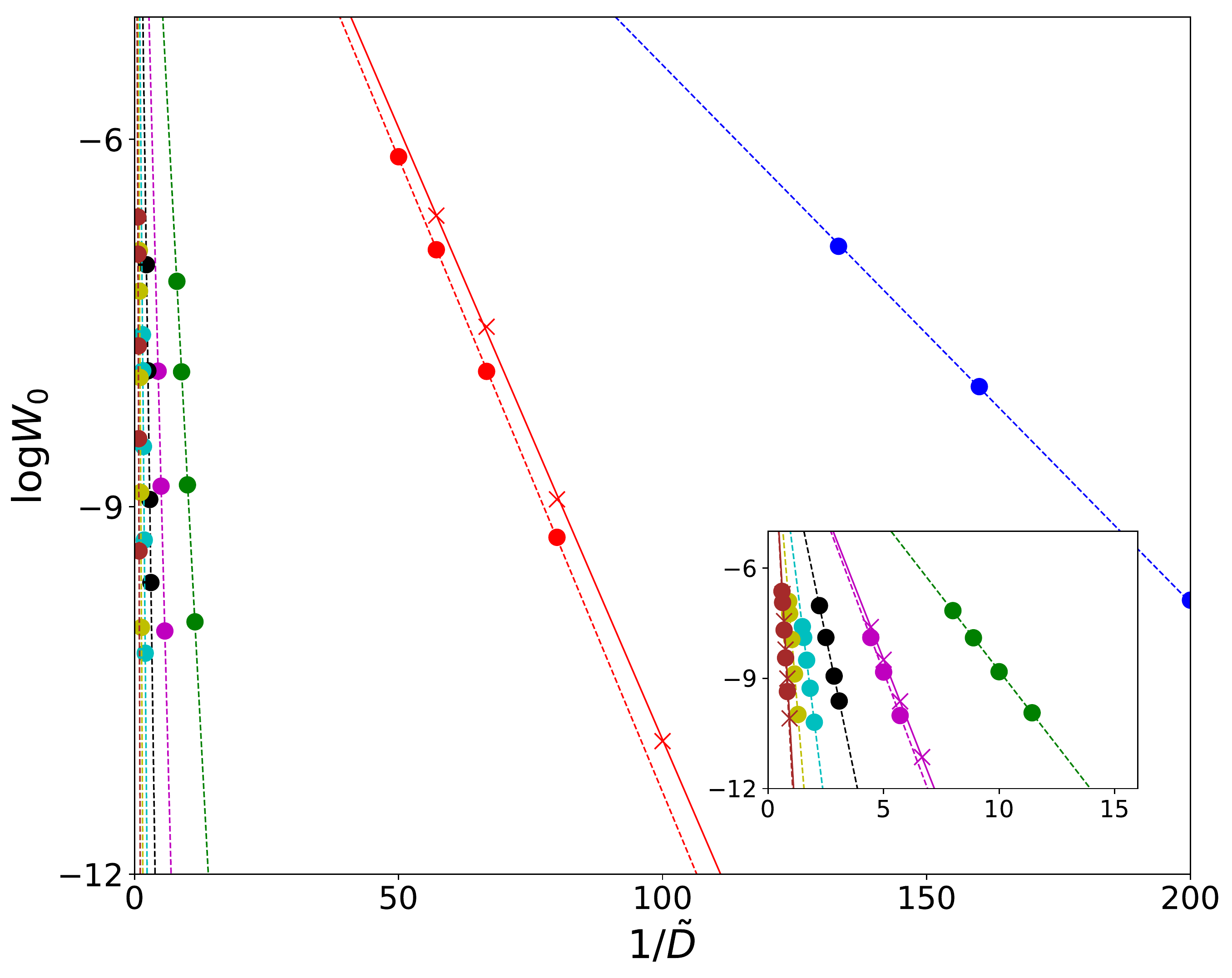}
\caption{The logarithm of the rate of escape from the zero-amplitude state  $\log W_0$ as a function of the scaled noise intensity $\tilde D=f^2D$ as obtained by numerically simulating the Langevin dynamics. The full circles show the results obtained from Eq.~(\ref{eq:eom_reduced}), where the driving field amplitude $f$ was scaled out. The crosses show the results obtained from the full equations (\ref{eq:eom})  for $f=5$. The results show excellent scaling $\log W_0\propto 1/\tilde D$. The straight dashed lines are guide for the eye. The data with brown, yellow, cyan, black, magenta, green, red, and blue full circles refer to $\kappa =4.5, 3.5, 3, 2.5, 2, 1.5, 0.5, 0.25$.}
\label{fig:simulations}
\end{figure}

In Fig~\ref{fig:simulations} we show the results of the numerical simulations. They clearly demonstrate the activation dependence of the escape rate on the noise intensity for all values of the scaled decay rate we explored: $\log W_0$ is linear in $1/\tilde D$. Already for $f=5$, the results obtained from the full equations (\ref{eq:eom}) were extremely close to those obtained from the scaled equations (\ref{eq:eom_reduced}) that refer to the $f\gg 1$ limit. 

\begin{figure}[t]
\includegraphics[scale=0.3]{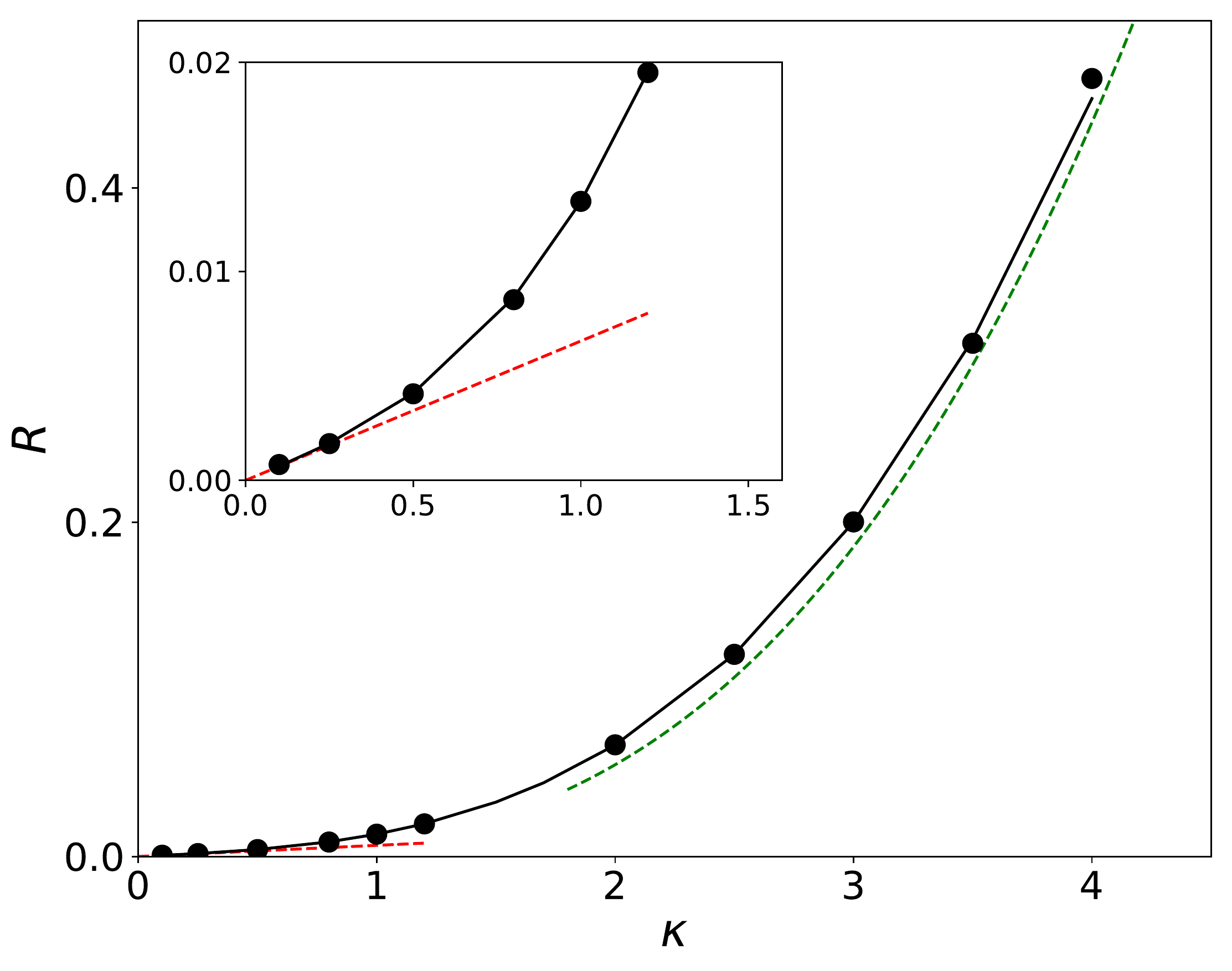}
\caption{The dependence of the effective activation energy $R$ of escape from the shallow metastable zero-amplitude state on the scaled decay rate of the mode $\kappa$. The results refer to strong driving, The solid lines show the solution of the full variational problem (\ref{eq:R_general}). The red and green dashed lines show the asymptotic results (\ref{eq:R_underdamped_explicit}) and (\ref{eq:large_kappa}) for small and large $\kappa$, respectively. The full circles show the values of $R$ obtained from numerical simulations of the stochastic equation of motion (\ref{eq:eom_reduced}).}
\label{fig:R_numerical}
\end{figure}

In Fig.~\ref{fig:R_numerical} we  compare the results of the simulations with the results obtained by solving the variational problems (\ref{eq:R_general}) and (\ref{eq:large_kappa}), se Appendix A. The effective activation energy of escape increases with the increasing decay rate $\kappa$. The results show that the effective activation energy $R$ is well-described by the asymptotic small-$\kappa$ expression (\ref{eq:R_underdamped_explicit}) in the range $\kappa\lesssim 0.5$. In the range $\kappa\gtrsim 1.5$ the asymptotic large-kappa expression (\ref{eq:R_large_kappa}) works reasonably well.

It is also seen from Fig.~\ref{fig:R_numerical} that the values of the activation energy obtained by simulations and from the analytical theory are in excellent agreement. This agreement holds for all values of $\kappa$ we explored.

\section{Conclusions}
\label{sec:conclusions}

The results of this paper bear on two problems. One is the possibility to facilitate the comparison of the theory and the experiment on escape from a metastable state of a generic thermally nonequilibrium system in the presence of weak noise. The second is the spontaneous breaking of the time-translation symmetry in vibrational systems where the symmetry-preserving state is dynamically stable. We have shown that both problems can be addressed by studying period tripling in mesoscopic vibrational systems, including nano-mechanical and Josephson-junction based systems in particular. Where such systems are driven close to triple the eigenfrequency, they can have three stable period-3 states and also a stable ``zero-amplitude'' state where they do not oscillate or oscillate with a small amplitude at the drive frequency.
 
The stability of the zero-amplitude state is symmetry-protected and, because of the symmetry, the dynamics in the vicinity of this state has no detailed balance. In this sense, this dynamics is generic for a nonequilibrium system. At the same time, escape from the zero-amplitude state into one of the period-3 states, and thus the breaking of the symmetry of the driving with respect to time translation by $2\pi/\omega_F$, can have a comparatively large probability for strong driving even where the noise is weak. This means that the zero-amplitude state becomes shallow while remaining dynamically stable. We have shown that the escape rate  $W_0$ displays universal scaling with the parameters of the vibrational mode as well as the amplitude $F$ and the frequency $\omega_F$ of the drive. For thermal noise Eq.~(\ref{eq:escape_general}) takes the form
\begin{align}
\label{eq:thermal_noise}
\log W_0 \approx -R(\kappa)/\tilde D, \qquad \tilde D=\frac{81}{64}\frac{\kappa F^2 k_BT}{\omega_F^4(\delta\omega)^2},
\end{align}
where $\kappa$ is the scaled decay rate. The dependence of the effective activation energy of escape $R$ on the single parameter $\kappa$ has been found analytically in the limiting cases and numerically in the general case. Quite remarkably, the simple explicit results in these limiting cases well describe the escape rate almost in the entire range of the parameters of the system.

The possibility to exponentially strongly increase the escape rate by varying the parameters of the driving is important for studying escape in the experiment. The explicit scaling of the escape rate with the parameters of the drive opens a viable approach to a detailed quantitative comparison of the theory of escape in systems lacking detailed balance with the experiment. 

\acknowledgements

M.I.D. acknowledges partial support from the NSF Grants No. DMR-1806473 and CMMI-1661618.

\appendix
\section{The real-time instanton}
A simple numerical  procedure of finding the MPEP is based on solving the Hamiltonian equations of motion for the auxiliary system with the Lagrangian ${\cal L}(d\qb/d\tau,\qb)$ given by Eq.~(\ref{eq:R_general}). The Hamiltonian of this system and the corresponding equations are 
 \begin{align}
 \label{eq:Hamiltonian_methods}
& {\cal H}(\qb,\pb) = \frac{1}{2}\pb^2 + \pb\kb, \qquad \frac{d\qb}{d\tau} = \pb + \kb,\qquad \frac{d\pb}{d\tau}= -\n(\pb\kb).
 \end{align}
The MPEP corresponds to the trajectory that starts for $\tau\to -\infty$ at the stable state of the original system and arrives for $\tau\to\infty$ to one of the saddle points, cf. \cite{Freidlin1998,Dykman1979b,Graham1984b,Chinarov1993,Maier1993a}. It is a real-time instanton. On this trajectory ${\cal H}=0$. The activation energy of escape $R$ is given by the dynamical action accumulated by moving along this trajectory,
\begin{align}
\label{eq:action_methods}
R=\frac{1}{2}S(\tau\to\infty), \qquad \frac{dS}{d\tau} = \pb\frac{d\qb}{d\tau}
\end{align} 
with $S(\tau\to-\infty)\to 0$.

The Hamiltonian equations (\ref{eq:Hamiltonian_methods}) can be solved by first linearizing them near the stationary point $\qb=\pb ={\bf 0}$; this  gives $\pb \approx \hat M\qb$, $S\approx \qb\hat M\qb /2$, and explicitly determines the matrix $\hat M$; we find $\hat M = 2\kappa\hat I$. One can then uses the shooting method by choosing a small $\qb(0)$, with $\pb(0), S(0)$ given by the above expression, and integrating the equations (\ref{eq:Hamiltonian_methods}) and  (\ref{eq:action_methods}) so that the trajectory approaches one of the three saddle points. An example of the MPEP obtained this way is shown in Fig.~\ref{fig:phase_portrait}~(b). The results were used to obtain the solid line in Fig.~\ref{fig:R_numerical}.

The green dashed line in Fig.~\ref{fig:R_numerical} was obtained by the same procedure, with the Hamiltonian of the form of ${\cal H}'(\qb',\pb') = \frac{1}{2}\pb'{}^2 +\pb'\kb'$, where the vector $\kb'\equiv \kb'(\qb') = (-q_1' +2q_1'q_2', -q_2' +q_1'{}^2 -q_2'{}^2)$.


%

\end{document}